\documentclass[11pt]{article}
\usepackage[margin=0.95 in]{geometry}
\usepackage{amsmath}
\usepackage{amssymb,amsfonts}
\usepackage[all]{xy}
\usepackage{graphicx}
\usepackage[utf8x]{inputenc}
\usepackage{amsmath}
\usepackage{amssymb}
\usepackage{float}
\usepackage{array}
\usepackage{tikz}
\usepackage{mathtools}
\usepackage{mathrsfs}
\usepackage{hyperref}
\usepackage{cite}
\numberwithin{equation}{section}
\numberwithin{equation}{section}
\numberwithin{table}{section}

\newcommand{\p}{\partial}
\newcommand{\be}{\begin{equation}}
\newcommand{\ee}{\end{equation}}

\begin{document}
\thispagestyle{empty}

\vspace*{3cm}
{}
\noindent
{\LARGE \bf  Modular properties of surface operators in $\mathcal N=2$ SQCD}
\vskip .4cm
\noindent
\linethickness{.06cm}
\line(10,0){467}
\vskip 1.1cm
\noindent
\noindent
{\large \bf Sourav Ballav$^a$ and Renjan Rajan John$^b$}
\vskip 0.5cm
{\em 
\noindent
$^a$Institute of Mathematical Sciences \\
Homi Bhabha National Institute (HBNI)\\
IV Cross Road, C.~I.~T.~Campus, \\
  Taramani, Chennai, 600113  Tamil Nadu, India}
\vskip 0.5cm
{\em
\noindent
$^b$Universit\`a di Torino, Dipartimento di Fisica\\
\hskip -.05cm
 I.\,N.\,F.\,N. - sezione di Torino, \\
Via P. Giuria 1, I-10125 Torino, Italy}
 \vskip 0.5cm
 {\em
 \noindent
sballav@imsc.res.in, renjan.rajan@to.infn.it
}
\vskip 1.2cm

\noindent {\sc Abstract: }  We study half-BPS surface operators in $\mathcal N=2$ supersymmetric QCD in four dimensions with gauge group SU(2) and four fundamental flavours. We compute the twisted chiral superpotential 
that describes the effective theory on the surface operator using equivariant localization as well as the Seiberg-Witten data. We then use the constraints imposed by S-duality to resum the instanton contributions to the twisted superpotential into elliptic functions and (quasi-) modular forms. The resummed results match what one would obtain from the description of surface operators as the insertion of a degenerate operator in a spherical conformal block in Liouville CFT.

\vskip1cm

\pagebreak

\newpage
\setcounter{tocdepth}{2}
\tableofcontents

\section{Introduction}
Half-BPS surface operators in $\mathcal N=4$ super Yang-Mills theories were introduced in \cite{Gukov:2006jk} as solutions to Hitchin equations with isolated singularities on a two-dimensional sub-manifold of the four dimensional space-time.
They are two dimensional generalizations of 't Hooft operators that provide details about the phase structure of the gauge theory. In the context of $\mathcal{N}=2$ theories they were introduced in \cite{Gukov:2007ck,Gaiotto:2009fs}.

From the point of view of four dimensional gauge theories obtained by wrapping M5-branes on Riemann surfaces with punctures \cite{Witten:1997sc,Gaiotto:2009we}, there are co-dimension 2 as well as co-dimension 4 surface defects. The former corresponds to the intersection of the original M5 branes with another stack of M5-branes and describes surface operators as singularities of the four dimensional fields on the two dimensional sub-manifold \cite{Alday:2010vg,Awata:2010bz,Kanno:2011fw}. The latter corresponds to M2 branes with boundaries on the M5 branes. In this description the surface operator is pointlike on the Riemann surface and the location labels the defect \cite{Alday:2009fs,KashaniPoor:2012wb,Kashani-Poor:2013oza,Kashani-Poor:2014mua,Gomis:2014eya,Gomis:2016ljm}.

In this paper we study surface operators in $\mathcal N=2$ SQCD with gauge group SU(2) and $N_f=4$ fundamental flavours. The matter content of the theory ensures that it is conformal in the limit that the flavour masses are zero. The low energy physics of the gauge theory on the Coulomb branch in the presence of the defect is described by two holomorphic functions, the prepotential and the twisted chiral superpotential. While the prepotential describes the effective four dimensional theory in the absence of a defect, the twisted chiral superpotential describes the effective theory on the defect. As a result, the twisted chiral superpotential is the quantity of interest to us in this paper.

We follow two approaches to compute the twisted chiral superpotential. The instanton moduli space in the background of a co-dimension 2 defect has been shown to be equivalent to the instanton moduli space on a suitable orbifold \cite{Awata:2010bz,Kanno:2011fw,Nawata:2014nca,Ashok:2017odt,Ashok:2017lko,Nekrasov:2017rqy,Jeong:2018qpc}. In Section \ref{section2} we use results from this approach to compute instanton corrections from localization methods. In Section \ref{section3} we follow \cite{Alday:2009fs} where it was proposed that the twisted chiral superpotential is obtained from the Seiberg-Witten data of the gauge theory. Although the proposal was for co-dimension 4 surface defects, for the SU(2) theory the two defects lead to the same IR behaviour as discussed in \cite{Frenkel:2015rda}. As a result we will not distinguish between the two in this work.
We compare results from the two approaches in the massless limit and obtain the map between parameters. The map ensures that results from the two approaches match when the masses are turned on. 

In Section \ref{section4} we come to the main topic of study in this paper. In \cite{Billo:2013fi,Billo:2013jba,Billo:2014bja,Billo:2015pjb,Billo:2015jyt}, S-duality was used to constrain the prepotential of $\mathcal N=2^\star$ theories with classical and exceptional gauge groups. The instanton expansion of the prepotential was resummed to a mass expansion such that the expansion coefficients were expressed as linear combinations of (quasi-) modular forms of the duality group. This was then done for asymptotically conformal SQCD with fundamental matter in \cite{Ashok:2015cba, Ashok:2016oyh}. This program was later extended to the case of a gauge theory with a surface defect in $\mathcal N=2^\star$ SU($N$) theory to constrain the twisted superpotential \cite{Ashok:2017odt}.
In the present work we extend this one step further by using S-duality to constrain the twisted superpotential of the SU(2) theory with $N_f=4$ fundamental flavours. The main difference from the $\mathcal N=2^\star$ theory is that both the gauge coupling as well as the continuous parameter that labels the surface defect get renormalized. As a result we require the map that relates the bare and the renormalized variables. We then solve the modular anomaly equation that the twisted chiral superpotential expressed in terms of renormalized variables satisfies at each order in a mass expansion. Combining results from localization we resum the instanton contributions at each order to specific linear combinations of elliptic functions and (quasi-) modular forms. 

We give some technical details on elliptic functions and modular forms in Appendix \ref{appmodular} and verify the map between the resummed and the bare variables in Appendix \ref{appb}.


\section{Surface operators as monodromy defects}
\label{section2}
In this section we study surface operators in $\mathcal{N}=2$ supersymmetric SQCD with gauge group SU($2$) and $N_f=4$ fundamental flavours in four dimensions as monodromy defects \cite{Gukov:2006jk}. In the SU($N$) gauge theory, such defects are classified by non-trivial partitions $\vec n$ of $N$. For the SU(2) theory, there is one monodromy defect that breaks the gauge group on the defect to the Levi subgroup :
\begin{align}
\label{Levi}
\mathbb L=U(1)\times U(1)
\end{align}
The defect also breaks the flavour symmetry to \cite{Ashok:2019rwa} :
\begin{align}
\label{flavourbreaking}
\mathbb F=S[U(2)\times U(2)]
\end{align}
The prepotential $\mathcal{F}$ and the twisted chiral superpotential $\mathcal{W}$ receive contributions from classical, 1-loop, and instanton terms :
\begin{align}
\label{noreno}
\mathcal{F}&=\mathcal{F}^\text{class}+\mathcal{F}^\text{1loop}+\mathcal{F}^{\text{inst}}\cr
\mathcal{W}&=\mathcal{W}^\text{class}+\mathcal{W}^\text{1loop}+\mathcal{W}^{\text{inst}}
\end{align}
The instanton contributions to $\mathcal F$ and $\mathcal W$ are obtained from the instanton partition function $\mathcal Z^{\text{inst}}$ as \cite{Alday:2009fs}
\begin{align} 
\label{extractfw}
\lim_{\epsilon_i\rightarrow 0}\log(1+\mathcal Z^{\text{inst}}[\vec n])=-\frac{\mathcal F^{\text{inst}}}{\epsilon_1\epsilon_2}+\frac{\mathcal W^{\text{inst}}}{\epsilon_1}
\end{align}
where $\epsilon_1$  and $\epsilon_2$ are the Omega-deformation parameters \cite{Nekrasov:2002qd,Nekrasov:2003rj}.
In the presence of a co-dimension 2 surface defect, $\mathcal Z^\text{inst}$ is obtained by the orbifold procedure detailed in \cite{Alday:2010vg,Awata:2010bz,Kanno:2011fw,Nawata:2014nca,Ashok:2017odt,Ashok:2017lko,Nekrasov:2017rqy,Jeong:2018qpc}. For the SU(2) theory with $N_f=4$, $\mathcal Z^{\text{inst}}$ is given by equations 8-9 of \cite{Ashok:2019rwa} with $M=2$ and $\vec n=[1,1]$ :
\begin{equation}
\label{Zso4d}
\mathcal Z^{\text{inst}}[1,1] =\sum_{\{d_1,d_2\}}\frac{(-q_1)^{d_1}}{d_1!}\frac{(-q_2)^{d_2}}{d_2!}\int\prod_{\sigma=1}^{d_1}\frac{d\chi_{1,\sigma}}{2\pi i}\,\int\prod_{\sigma=1}^{d_2}\frac{d\chi_{2,\sigma}}{2\pi i}z_{\{d_1,d_2\}}\,,
\end{equation}
where $q_1$ and $q_2$ are the instanton counting parameters, $d_1$ and $d_2$ the number of ramified instantons, $\hat\epsilon_2\equiv\frac{\epsilon_2}{2}$, $m_1,\ldots, m_4$ the masses of fundamental flavours, and
\begin{align}
\label{zexplicit4d}
z_{\{d_1,d_2\}}&= \prod_{\sigma,\tau=1}^{d_1}\frac{\left(\chi_{1,\sigma} - \chi_{1,\tau} + \delta_{\sigma,\tau}\right)}{\left(\chi_{1,\sigma} - \chi_{1,\tau} +\epsilon_1\right)}  \prod_{\sigma,\tau=1}^{d_2}\frac{\left(\chi_{2,\sigma} - \chi_{2,\tau} + \delta_{\sigma,\tau}\right)}{\left(\chi_{2,\sigma} - \chi_{2,\tau} +\epsilon_1\right)}\cr
&\prod_{\sigma=1}^{d_1}\prod_{\rho=1}^{d_{2}}\,
\frac{\left(\chi_{1,\sigma} - \chi_{2,\rho} + \epsilon_1 + \hat\epsilon_2\right)}
{\left(\chi_{1,\sigma} - \chi_{2,\rho} + \hat\epsilon_2\right)} \prod_{\sigma=1}^{d_2}\prod_{\rho=1}^{d_{1}}\,
\frac{\left(\chi_{2,\sigma} - \chi_{1,\rho} + \epsilon_1 + \hat\epsilon_2\right)}
{\left(\chi_{2,\sigma} - \chi_{1,\rho} + \hat\epsilon_2\right)}\cr
&\prod_{\sigma=1}^{d_1} \frac{(\chi_{1,\sigma} - m_{1})(\chi_{1,\sigma} - m_{2})}
{\left(a_{1}-\chi_{1,\sigma} + \frac 12 (\epsilon_1 + \hat\epsilon_2)\right)
\left(\chi_{1,\sigma} - a_{2} + \frac 12 (\epsilon_1 + \hat\epsilon_2)\right)}\cr
&\prod_{\sigma=1}^{d_2} \frac{(\chi_{2,\sigma} - m_{3})(\chi_{2,\sigma} - m_{4})}
{\left(a_{2}-\chi_{2,\sigma} + \frac 12 (\epsilon_1 + \hat\epsilon_2)\right)
\left(\chi_{2,\sigma} - a_{1} + \frac 12 (\epsilon_1 + \hat\epsilon_2)\right)}\,\,.
\end{align}
Here $a_1$ and $a_2$ are the Coulomb vev's and upon imposing the SU(2) constraint we have $a_1=-a_2=a$.
Since the integral in \eqref{Zso4d} is a contour integral, it requires us to prescribe a contour of integration to pick the poles that contribute. The various allowed prescriptions are captured by what is called the Jeffrey-Kirwan (JK) reference vector \cite{JK} and 
it was shown in \cite{Gorsky:2017hro,Ashok:2018zxp,Ashok:2019rwa} that different contour prescriptions map to Seiberg dual descriptions of surface operators as 2d/4d coupled systems.
Our choice of contour is such that the integral picks the poles in the upper half $\chi_{1,2}$ plane \cite{Gorsky:2017hro,Ashok:2017lko}.
%
We package the instanton contributions to $\mathcal F$ and $\mathcal W$ as \footnote{When we package the entire prepotential or the twisted chiral superpotential as in \eqref{MassExp} we use no superscript.} 
\begin{align}
\label{MassExp}
\mathcal{F}^\text{inst}=\sum_{n=0}^{\infty} f_n^\text{inst},\quad\quad
\mathcal{W}^\text{inst}= \sum_{n=0}^{\infty} w_n^\text{inst}
\end{align}
where $f_{n}^\text{inst}\sim a^{2-n}$ and $w_n^\text{inst}\sim a^{1-n}$. From \eqref{Zso4d} and \eqref{extractfw} one obtains :
\begin{align}
f_{2k+1}^{\text {inst}}&=0,\quad\forall\quad k\in\mathbb Z_{\ge 0}\cr
w_{2k+1}^{\text {inst}}&=0,\quad\forall\quad k\in\mathbb Z^{+}\,.
\end{align}
The first few non-zero $f_n^\text{inst}$ up to 4 ramified instantons are :
\begin{align}
\label{masslessf}
f_0^\text{inst}&=a^2\left[\frac{q_1 q_2}{2}+\frac{13(q_1 q_2)^2}{64}\right]\cr
f_2^\text{inst}&=\frac{q_1 q_2}{2}\sum_{i<j}m_im_j+\frac{(q_1 q_2)^2}{64}\left(\sum_{i}m_i^2+16\sum_{i<j}m_im_j\right)\cr
f_4^\text{inst}&=\frac{1}{2a^2}\left[q_1 q_2m_1m_2m_3m_4+\frac{(q_1 q_2)^2}{32}\left(16m_1m_2m_3m_4+\sum_{i<j}m_i^2m_j^2\right)\right]\,.
\end{align}
We now give the first few non-zero $w_n^\text{inst}$ up to 4 ramified instantons :
\begin{align}
\label{masslessw}
w_0^\text{inst}&=a\Bigg[\frac{q_1}{2}+\frac{3q_1^2}{16}+\frac{5q_1^3}{48}+\frac{35q_1^4}{512}-(q_1\rightarrow q_2)+\frac{q_1q_2}{16}\left(q_1+\frac{q_1^2}{2}-(q_1\rightarrow q_2)\right)\Bigg]\cr
w_{1}^\text{inst}&=-\frac{m_1+m_2}{2}\left(q_1+\frac{q_1^2}{2}+\frac{q_1^3}{3}+\frac{q_1^4}{4}\right)-\left(m_{1,2}\rightarrow m_{3,4},q_1\rightarrow q_2\right)\cr
w_{2}^\text{inst}&=\frac{1}{a}\Bigg[\frac{\left(m_1^2+m_2^2\right)}{16}\left(q_1^2+q_1^3+\frac{15 }{16}q_1^4-q_1^2 q_2+\frac{q_1^2 q_2^2}{8}-\frac{q_1^3 q_2}{2}\right)
-\Big(m_{1,2}\rightarrow m_{3,4},q_1\leftrightarrow q_2\Big)\cr
&\hspace{1cm}+\frac{m_1m_2}{2}\left(q_1+\frac{q_1^2}{2}+\frac{3q_1^3}{8}+\frac{5q_1^4}{16}-\frac{q_1q_2}{2}-\frac{q_1q_2^2}{8}-\frac{q_1q_2^3}{16}-\frac{3}{16}q_1^2q_2^2-\frac{q_1^3q_2}{16}\right)\cr
&\hspace{3cm}-\Big(m_{1,2}\rightarrow m_{3,4},q_1\leftrightarrow q_2\Big)\Bigg]\cr
w_{4}^\text{inst}&=-\frac{1}{16 a^3}\Bigg[\frac{1}{32}\left(m_1^4 +m_2^4\right)q_1^4 -\left(m_3^4+m_4^4\right)q_2^4+\frac{m_1 m_2 m_3 m_4}{2}\left(q_1^3 q_2-q_1 q_2^3\right)\cr
&\hspace{2cm}+\left(m_1^3 m_2+m_1 m_2^3\right)\left(\frac{q_1^3}{3}-\frac{q_1^3 q_2}{2}+\frac{q_1^4}{2}\right)
-\left(m_{1,2}\rightarrow m_{3,4},q_1\leftrightarrow q_2\right)\cr
&\hspace{2cm}+m_1^2 m_2^2\left(q_1^2+q_1^3+\frac{9}{8}q_1^4-q_1^2 q_2+\frac{q_1^2 q_2^2}{4} -\frac{q_1^3 q_2}{2} \right)
-\left(m_{1,2}\rightarrow m_{3,4},q_1\leftrightarrow q_2\right)\cr
&\hspace{2cm}+\left(m_1^2 +m_2^2\right)m_3 m_4\left(q_1^2 q_2-\frac{q_1^2 q_2^2}{2} +\frac{q_1^3 q_2}{2} \right)-\left(m_{1,2}\leftrightarrow m_{3,4},q_1\leftrightarrow q_2\right)\Bigg]
\end{align}
where we have used $(\rightarrow,\leftrightarrow)$ to denote terms that are obtained by performing the switch indicated by the arrows on the immediately preceding terms.
%
In order to confirm the above results for the twisted superpotential obtained via localization, we will now compute the same from the Seiberg-Witten (SW) data of the gauge theory. 
%
\section{Superpotential from Seiberg-Witten data}
\label{section3}
In this section we follow the proposal in \cite{Alday:2009fs} according to which the twisted chiral superpotential can be computed from the SW data. This helps us obtain the map that relates the gauge theory parameters to the instanton counting parameters and thereby verify the results from localization obtained in the previous section.
According to the proposal in  \cite{Alday:2009fs} the twisted superpotential is given by the integral of the SW differential $\lambda$ along an open path on the SW curve :
\begin{equation}
\label{GGSproposal}
{\mathcal W}(x_0) = \int^{x_0} \lambda\,,
\end{equation}
where $x_0$ is the continuous parameter that labels the surface operator, and is given by the location of the defect on the Riemann surface.

Let us now recall a few salient features of the SW solution of the SU$(2)$ theory with $N_f=4$ flavours. We will work with the Gaiotto form of the curve as $\lambda$ is easily extracted from there.
The Gaiotto form of the curve is \cite{Gaiotto:2009we}
\be
\label{Gcurvegeneral}
x^2 = \phi_2(t) \,,
\ee
where $\phi_2(t)dt^2$ is a quadratic differential.  
The SW differential is readily given by \cite{Gaiotto:2009we}
\be
\label{Gdifffgeneral}
\lambda = x\, dt = \sqrt{\phi_2(t)} dt \,.
\ee
%
Let us first analyse the case when the masses of the flavours are set to zero. In this limit, the Gaiotto curve is such that $ \phi_2(t)$ takes the form \cite{Ashok:2015gfa}
\be
\label{quaddiff}
\phi_2(t)=\frac{q_0(q_0-1)}{t(t-q_0)(t-1)}\frac{\partial f_0}{\partial q_0}\,.
\ee
where $q_0=e^{\pi i\tau_0}$, such that  $\tau_0=\frac{\theta}{\pi}+\frac{8\pi i}{g^2}$ is the bare complexified gauge coupling and $f_0$ is the prepotential in the massless limit. After adding the classical and the 1 loop terms to the instanton contribution obtained via the SW analysis (see \cite{Ashok:2015gfa} for example) which matches the localization results obtained in the previous section we have  :
\begin{align}
\label{f0masslessq}
f_0&=a^2\log q_0-a^2\log 16+f_0^\text{inst}\cr
&=a^2\left(\log q_0-\log 16+\frac{q_0}{2}+\frac{13q_0^2}{64}\right)\,.
\end{align}
We substitute for the SW differential from \eqref{Gdifffgeneral}, \eqref{quaddiff} and \eqref{f0masslessq}, and perform the integral in \eqref{GGSproposal} to obtain :
\begin{align}
\label{WfromAGGTV}
w_0=a\log x_0+a\Bigg[\frac{x_0}{2}+\frac{3x_0^2}{16}+\frac{5x_0^3}{48}+\frac{35x_0^4}{512}-\left(x_0\rightarrow \frac{q_0}{x_0}\right)+\frac{q_0}{16}\left(x_0+\frac{x_0^2}{2}-\left(x_0\rightarrow\frac{q_0}{x_0}\right)\right)\Bigg]\,.
\end{align}
By comparing $w_0^{\text{inst}}$ from \eqref{masslessw} and $w_0$ obtained from the curve \eqref{WfromAGGTV}, we obtain the following map between the instanton counting parameters $(q_1,q_2)$ and the gauge theory parameters $(q_0,x_0)$ :
\begin{align}
\label{paramap}
q_1=x_0,\quad q_2=\frac{q_0}{x_0}\,.
\end{align}
Note that in $f_n^\text{inst}$ in \eqref{masslessf} $q_1$ and $q_2$ always appear as the combination $q_1q_2$ and powers thereof, thus ensuring that the prepotential depends only on $q_0$ and is independent of $x_0$.
%

We will now consider the case when all the masses are turned on. The Gaiotto form of the SW curve is still $x^2=\phi_2(t)$, where \cite{Ashok:2015gfa}:
\begin{align}
\label{GaiottoMass}
 \phi_2(t)&=\frac{q_0(q_0-1)}{t(t-1)(t-q_0)}\,\frac{\p\mathcal F}{\p q_0}+\frac{q_0(m_1+m_2)(m_3+m_4)}{2t(t-1)(t-q_0)}-\frac{(q_0-1)(m_3^2+m_4^2)}{2t(t-1)(t-q_0)}\cr
&\hspace{.5cm}-\frac{m_3^2+m_4^2+2m_1m_2}{2t(t-1)}+\frac{(m_3-m_4)^2}{4t^2}+\frac{(m_3+m_4)^2}{4(t-q_0)^2}+\frac{(m_1+m_2)^2}{4(t-1)^2}\,.
\end{align}
The twisted superpotential when the masses are turned on is obtained exactly as in the massless case by performing the integral in \eqref{GGSproposal}. 
One can easily check that the instanton contributions to $\mathcal W$ obtained via localization in the previous section matches the results from the SW data after the masses are turned on, provided one uses the map \eqref{paramap} between parameters.
We have checked that the match holds up to $w_8$ to 8 ramified instantons. 

Now that we have matched $\mathcal W$ obtained via localization and from the SW data we will shift gears and turn our attention to utilizing the S-duality symmetry of the theory to resum the instanton contributions.
%
%
\section{Resumming the twisted chiral superpotential}
\label{section4}
As mentioned in the Introduction, a lot of progress has been made in resumming the instanton contribution to the prepotential of a large class of theories into (quasi-) modular forms of their respective S-duality groups \cite{Billo:2013fi,Billo:2013jba,Billo:2014bja,Billo:2015pjb,Billo:2015jyt,Ashok:2015cba, Ashok:2016oyh}. This was then extended to the case of the twisted chiral superpotential of $\mathcal N=2^\star$ SU($N$) theory in the presence of a surface defect in \cite{Ashok:2017odt}. There it was shown that $\mathcal W$ satisfies a modular anomaly equation, 
and that the instanton expansion of $\mathcal W$ at each order in a mass expansion can be resummed into elliptic functions and (quasi-) modular forms. 
Since the SU(2) theory with $N_f=4$  also has an S-duality symmetry we will now attempt to do the same in this theory. 
\subsection{Resummation variables}
Unlike in the $\mathcal N=2^\star$ theory, the gauge coupling and the continuous parameter that describes the surface defect are renormalized in asymptotically conformal SQCD theories. In the theory of interest to us  this is already clear from the expressions for $\mathcal F$ and $\mathcal W$ in the massless limit in \eqref{f0masslessq} and \eqref{WfromAGGTV} respectively. We would like to resum the terms on the RHS of these equations to simple expressions in terms of the renormalized counterparts $q$ and $x$ of $q_0$ and $x_0$ respectively.
 The $q_0$ vs $q$ relation has appeared in several references and is given by \cite{Grimm:2007tm,Billo:2013fi,Kashani-Poor:2013oza,Ashok:2016oyh}  :
\begin{align}
\label{renocoupling}
q_0=\frac{e_3-e_2}{e_1-e_2}(q)=\frac{\theta_2^4(q)}{\theta_3^4(q)}
\end{align}
where $e_i\equiv\wp(\omega_i)$ denote the Weierstra\ss\, $\wp$ function evaluated at the half periods and $\theta_i$ are the Jacobi $\theta$ functions. 
We refer the reader to Appendix \ref{appmodular} for details on Jacobi theta functions and the Weierstra\ss\, $\wp$ function.
The first few terms that appear in the expansion of \eqref{renocoupling} are :
\begin{align}
\label{qexp}
q_0&=16q(1-8q+44q^2-192q^3+\ldots)
\end{align}
One can now check that $\widetilde f_0$ which is the prepotential in the massless limit \eqref{f0masslessq} when expressed in terms of $q$ takes the expected form :
\begin{align}
\label{fotilde}
\widetilde f_0=a^2\log q\,.
\end{align}
Note that here and henceforth we use the tilde symbol to denote quantities expressed in terms of the renormalized variables $(q,x)$. 

For the parameter $x_0$, following the analysis in \cite{Kashani-Poor:2013oza} we propose the following map to the resummed variable $x$  : 
\be
\label{x0map}
x_0=\frac{\wp(z+w_1|\,\tau)- e_2}{e_1- e_2} 
\ee
where $\tau$ and $z$ are such that 
\begin{align}
\label{newvars}
q=\exp(\pi i\tau),\quad x=\exp(2\pi i z)\,. 
\end{align}
A similar map was also used in the recent paper \cite{He:2019dso}. We verify this map in Appendix \ref{appb} using the SW analysis in the massless limit. Note that the $q_0$ vs $q$ map in \eqref{renocoupling} is a special case of \eqref{x0map} for $z=w_2$.
The first few terms that appear in the expansion of \eqref{x0map} are :
\begin{align}
\label{xexp}
x_0&=4(x-2x^2+3x^3-4x^4)+8q(1-4x+8x^2-12x^3)+4q^2\left(\frac{1}{x}-12\right)+\ldots
\end{align}
With the above expansions for $q_0$ and $x_0$ one can check that up to purely $q_0$ dependent terms, $\widetilde w_0(q,x)$ which is the twisted superpotential in the massless limit \eqref{WfromAGGTV} takes the expected form :
\begin{align}
\label{wotilde}
\widetilde w_0=a\log x\,.
\end{align}
In the next section where we resum the instanton contributions to $\mathcal W$ we will find it more convenient to work with the $\log x$ derivative of $\widetilde{\mathcal W}$ whose expansion is :
\begin{align}
\label{wprimeexpreno}
x\frac{\partial\mathcal {\widetilde W}}{\partial x}\equiv\mathcal{\widetilde W'}
=\sum_{n=0}^\infty \widetilde w_n'
\end{align}
where $\widetilde w_n'\sim a^{1-n}$. Clearly from \eqref{wotilde} we have : 
\begin{align}
\label{zeromasswoprime}
\widetilde w_0'&=a\,.
\end{align}
The next few non-zero $\widetilde w_n'$ obtained by substituting the expansions for $q_0$ and $x_0$ from \eqref{qexp} and \eqref{xexp} in \eqref{masslessw} are :
\begin{align}
\label{renormalizedvars}
\widetilde w_1'&=-2(m_1+m_2)\left(x+x^3-\frac{q^2}{x}\right)-2(m_3+m_4)\left(q\,x-\frac{q}{x}\right)\cr
\widetilde w_2'&=\frac{1}{a}\Big[2(m_1^2+m_2^2)(x^2+2x^4)+\frac{2q^2}{x^2}(m_3^2+m_4^2)+2m_1m_2\left(x+3x^3+\frac{q^2}{x}\right)\cr
&\hspace{4.5cm}+2m_3m_4\left(\frac{q}{x}+q\,x\right)\Big]\cr
\widetilde w_4'&=-\frac{1}{a^3}\Big[2(m_1^4+m_2^4)x^4+4m_1m_2(m_1^2+m_2^2)x^3+2m_1^2m_2^2(x^2+8x^4)+2m_3^2m_4^2\,\frac{q^2}{x^2}\cr
&\hspace{1.5cm}+4m_3m_4(m_1^2+m_2^2)q\,x+4m_1m_2(m_3^2+m_4^2)\frac{q^2}{x}+16m_1m_2m_3m_4\,q\,x^2\Big]
\end{align}
The above expressions will be useful in the next sub-section when we resum $\widetilde w_n'$ to linear combinations of elliptic functions and (quasi-) modular forms. 
\subsection{Modular Anomaly Equation for the twisted superpotential}
It is well known from \cite{Seiberg:1994aj} that the SU(2) theory with $N_f=4$ enjoys an S-duality symmetry under which the renormalized gauge coupling $\tau$ transforms as 
\begin{align}
\tau\rightarrow -\frac{1}{\tau}\,.
\end{align}
It was shown in \cite{Gukov:2006jk} (see also \cite{Ashok:2017odt}) that under this duality the variable $z$ that is related to the continuous parameter $x$ that labels the defect as in \eqref{newvars} transforms as
\begin{align}
z\rightarrow -\frac{z}{\tau}\,.
\end{align}
The action of S-duality on the Coulomb vev $a$ is such that%
\begin{align}
\label{Sduala}
S(a):=a_D=\frac{1}{2\pi i}\frac{\partial\mathcal F}{\partial a}=\tau\left(a+\frac{\delta}{12}\frac{\partial f}{\partial a}\right)
\end{align}
where $\delta=\frac{6}{\pi i\tau}$ and $f=\mathcal F^{\text{1 loop}}+\mathcal F^{\text{inst}}$. 
The anomalous terms on the RHS arise solely from the dependence of the prepotential on the second Eisenstein series $E_2$ \cite{Billo:2013fi}. From the form of $\widetilde w_0'$ in \eqref{zeromasswoprime}, we see that it transforms exactly as in \eqref{Sduala}.

\noindent Motivated by the transformation of $\widetilde{\mathcal W'}^{\text{class}}$ we propose that, as in \cite{Ashok:2017odt}, $\widetilde{\mathcal W'}$ transforms under S-duality with weight one.
%
The $\widetilde w_n'$ in \eqref{wprimeexpreno} then obey a modular anomaly equation, the derivation of which proceeds exactly as in the case of the $\mathcal N=2^\star$ theory in \cite{Ashok:2017odt}. The anomaly equation is :
\begin{align}
\label{MAE}
\frac{\partial \widetilde w_n'}{\partial E_2}+\frac{1}{12}\sum_{l=0}^{n-1}\left(\frac{\partial\widetilde w'_\ell}{\partial a}\right)\left(\frac{\partial\widetilde f_{n-\ell}}{\partial a}\right)=0 
\end{align}
Since $\widetilde w_1$ and $\widetilde w_1'$ are independent of $a$, they do not contribute to the IR dynamics and we start our analysis at $n=2$. For $n=2$ the equation takes the form :
\begin{align}
\label{level1}
\frac{\partial \widetilde w_2'}{\partial E_2}+\frac{1}{12}\left(\frac{\partial \widetilde w'_0}{\partial a}\right)\left(\frac{\partial \widetilde f_2}{\partial a}\right)=0 
\end{align}
The prepotential for this theory was resummed in \cite{Billo:2013fi} and in particular :
\begin{align}
\label{f2reno}
\widetilde f_2=2R\log\left(\frac{a}{\Lambda}\right)\,.
\end{align}
where 
\begin{align}
R&=\frac{1}{2}\sum_{f=1}^4 m_f^2\,.
\end{align}
We substitute for $\widetilde w'_0$ from \eqref{zeromasswoprime} and $\widetilde f_2$ from \eqref{f2reno} and solve \eqref{level1} to obtain,
\begin{align}
\label{w2resum}
\widetilde w_2'&=-\frac{E_2R}{6a}+\frac{1}{a}\,(\text{modular\ term}) 
\end{align}
Since the modular terms that one must add to \eqref{w2resum} must have weight two, one arrives at the following ansatz for $\widetilde w_2'$ : 
\begin{align}
\label{ansatzw2prime}
\widetilde w_2'&=-\frac{E_2R}{6a}+\frac{1}{a}\sum_{A=0}^3c_A\,\wp(z+\omega_A)
\end{align}
The coefficients $c_A$ are fixed by comparing the expansion of the RHS of the above equation with the first few terms in the localization result for the same expressed in terms of $(q,x)$ in \eqref{renormalizedvars}. 
This leads to : 
\begin{align}
\label{w2final}
\widetilde w_2'=-\frac{1}{6a}\sum_{A=0}^3M_A^2\left(E_2+12\widehat\wp(z+\omega_A)\right)
\end{align}
where $M_A$ are the following mass combinations :
\begin{align}
\label{bigmasses}
M_0=-\frac{(m_1+m_2)}{2},\,\, M_1=\frac{(m_1-m_2)}{2},\,\, M_2=\frac{(m_3+m_4)}{2},\,\, M_3=\frac{(m_3-m_4)}{2}\,,
\end{align}
which appear as residues of the quadratic differential in the SW data.
From the resummed result for $w_2'$ in \eqref{w2final}, one can see that under the combined action of S-duality on the gauge coupling and the triality transformation on the masses of the fundamental flavours, the $a$ independent part transforms as a quasi-modular form of weight two.

We performed a similar analysis of \eqref{MAE} at the next two levels. This required the following resummed expressions for the prepotential at $n=4,6$ \cite{Billo:2013fi} :
\begin{align}
\widetilde f_4&=-\frac{R^2E_2}{6}+T_1\theta_4^4-T_2\theta_2^4\cr
\widetilde f_6&=-\frac{R^3\left(5E_2^2+E_4\right)}{180a^4}-\frac{N E_4}{5a^4}\cr
&\hspace{.5cm}+\frac{RT_1\theta_4^4\left(2E_2+2\theta_2^4+\theta_4^4\right)}{6a^4}-\frac{RT_2\theta_2^4\left(2E_2-2\theta_4^4-\theta_2^4\right)}{6a^4}
\end{align}
where 
	\begin{align}
T_1&=\frac{1}{12}\sum_{f<f'=1}^4m_f^2m_{f'}^2-\frac{1}{24}\sum_{f=1}^4 m_f^4\cr
T_2&=-\frac{1}{24}\sum_{f<f'=1}^4m_f^2m_{f'}^2+\frac{1}{48}\sum_{f=1}^4 m_f^4-\frac{1}{2}m_1m_2m_3m_4\cr
N&=\frac{3}{16}\sum_{f<f'<f''=1}^4m_f^2m_{f'}^2m_{f''}^2-\frac{1}{96}\sum_{f\ne f'=1}^4m_f^2m_{f'}^4+\frac{1}{96}\sum_{f=1}^4m_f^6\,.
\end{align}
Here $R$, $T_i$, and $N$ are the first few mass invariants that transform under the triality action as :
\begin{align}
R\rightarrow R,\quad T_1\leftrightarrow T_2,N\rightarrow N\,.
\end{align}
Solving the modular anomaly equation \eqref{MAE} at $n=4,6$ we obtained the following resummed results for $\widetilde w_4'$ and $\widetilde w_6'$ :
\begin{align}
\label{w4primeresummed}
\widetilde w_4'&=-\frac{1}{72a^3}\Big(\sum_{A=0}^3M_A^4\left(2E_2^2-E_4+24E_2\widehat\wp(z+\omega_A)+144\widehat\wp(z+\omega_A)^2\right)\cr
&\hspace{1cm}+2\sum_{A<B}M_A^2M_B^2\Big(2E_2^2-E_4+12E_2\widehat\wp(z+\omega_A)+12E_2\widehat\wp(z+w_B)\cr
&\hspace{1cm}+144\widehat\wp(z+\omega_A)\widehat\wp(z+\omega_B)\Big)-12T_1\theta_4^4(E_2-2\theta_2^4-\theta_4^4)+12T_2\theta_2^4(E_2+\theta_2^4+2\theta_4^4)\Big)\cr
%
%
		\widetilde w_6'&=-\frac{1}{432a^5}\left(\sum_{A=0}^3M_A^2\left(E_2+12\widehat\wp\left(z+\omega_A\right)\right)\right)\Bigg(\sum_{B=0}^3M_B^4\Big(2E_2^2-E_4+24E_2\widehat\wp(z+\omega_B)\cr
&\hspace{.5cm}+144\widehat\wp(z+\omega_B)^2\Big)+2\sum_{B<C}M_B^2M_C^2\Big(2E_2^2-E_4+12E_2\widehat\wp(z+\omega_B)+12E_2\widehat\wp(z+w_C)\cr
&\hspace{0.5cm}+144\widehat\wp(z+\omega_B)\widehat\wp(z+\omega_C)\Big)-12T_1\theta_4^4(E_2-2\theta_2^4-\theta_4^4)+12T_2\theta_2^4(E_2+\theta_2^4+2\theta_4^4)\Bigg)\cr
&\hspace{.5cm}-\frac{R^3}{720a^5}(5E_2^3-E_2E_4-2E_6)
		-\frac{N}{15a^5}(E_2E_4-E_6)+\frac{R}{12a^5}(T_1\theta_4^4-T_2\theta_2^4)(E_2^2-E_4)
	\end{align}
%
%
%
%
%
Note that as in the case of $\widetilde w_2'$, under the combined action of S-duality and triality, $\widetilde w_4'$ and $\widetilde w_6'$ transform as expected. The above resummed results have been matched with explicit results from localization expressed in terms of the renormalized variables $(q,x)$ up to 8 ramified instantons.
%
\section{Summary}
\label{CFTnvd}

In this paper, we considered surface defects in SU(2) theory with four fundamental flavours and studied the twisted chiral superpotential as an expansion in the masses. We matched the results for the superpotential obtained from localization methods and from the Seiberg-Witten data. The coefficients in the mass expansion satisfy a modular anomaly equation that allows one to solve for them in an iterative manner in terms of (quasi-) modular and elliptic functions. A key input here is the explicit localization results that are crucial to fix the purely modular and elliptic contributions. 
While such an equation was known for the $\mathcal N=2^\star$ theory, the main difference now is that the variables in terms of which the resummation is done are not the bare couplings but the renormalized ones. This required us to write down the map that relates the bare and the renormalized variables. The map is verified using the Seiberg-Witten analysis in Appendix \ref{appb}.
%
%

In \cite{Alday:2009fs} it was shown that for the SU(2) $N_f=4$ theory the instanton partition function in the presence of the defect is reproduced by a 4-point spherical conformal block in Liouville CFT with the insertion of a degenerate primary. 
This was studied in great detail in \cite{Kashani-Poor:2013oza} and we have checked up to $n=6$ that our resummed results for $\widetilde w_n'$ match the results one would obtain following the CFT analysis in \cite{Kashani-Poor:2013oza}.
\section*{Acknowledgments}
We thank Marco Billo, Alberto Lerda, Prashanth Raman, and especially Sujay K. Ashok and Marialuisa Frau for many discussions and for helpful comments on an earlier version of the manuscript. The work of R.R.J is partially supported by the MIUR PRIN Contract 
2015MP2CX4 ``Non-perturbative Aspects Of Gauge Theories And Strings''. 
\appendix
\section{Useful formulas for modular forms and elliptic functions}
\label{appmodular}
The Jacobi $\theta$-functions are 
\begin{align}
\label{JacobiTheta}
\theta_1(z|\tau)&=\sum_{n=-\infty}^{\infty}\,q^{\left(n-\frac{1}{2}\right)^2}(-x)^{n-\frac{1}{2}}\cr
\theta_2(z|\tau)&=\sum_{n=-\infty}^{\infty}\,q^{\left(n-\frac{1}{2}\right)^2}x^{n-\frac{1}{2}}\cr
\theta_3(z|\tau)&=\sum_{n=-\infty}^{\infty}\,q^{n^2}x^n\cr
\theta_4(z|\tau)&=\sum_{n=-\infty}^{\infty}\,q^{n^2}(-x)^n
\end{align}
where $x = e^{2\pi i z}$ and $q=e^{\pi i\tau}$. At $z=0$, $\theta_2$, $\theta_3$ and $\theta_4$ give the following expansions
\begin{align}
\theta_2(0|\tau)\equiv\theta_2(q)&=2q^{1/4}(1+q^2+q^6+\ldots)\cr
\theta_3(0|\tau)\equiv\theta_3(q)&=1+2q+2q^4+2q^9+\ldots\cr
\theta_4(0|\tau)\equiv\theta_4(q)&=1-2q+2q^4-2q^9+\ldots
\end{align}
Under $\tau\rightarrow\tau'=-\frac{1}{\tau}$ these transform as follows :
\begin{align}
\theta_2^4\rightarrow -\tau^2\theta_4^4,\quad\theta_3^4=-\tau^2\theta_3^4,\quad\theta_4^4=-\tau^2\theta_2^4
\end{align}
The expansions to the first few orders of the first three Eisenstein series are given by
\begin{align}
E_2&=1-24q^2-72q^4+\ldots\cr
E_4&=1+240q^2+2160q^4+\ldots\cr
E_6&=1-504q^2-16632q^4+\ldots
\end{align}
While $E_4(\tau)$ and $E_6(\tau)$ transform as modular forms with weight 4 and 6 respectively, $E_2(\tau)$ is quasi-modular of degree 2.
Under $\tau\rightarrow\tau'=-\frac{1}{\tau}$ we have the following transformations :
\begin{align}
E_2(\tau')&=\tau^2E_2(\tau)+\frac{6}{i\pi}\tau\cr
E_4(\tau')&=\tau^4E_4(\tau)\cr
E_6(\tau')&=\tau^6E_6(\tau)
\end{align}
The Weierstra\ss~$\wp$-function is defined as
\begin{equation}
\label{wpdefn0}
 \wp(z|\tau)=-\frac{\partial^2}{\partial z^2}\log\theta_1(z|\tau)-\frac{\pi^2}{3}E_2(\tau) ~.
\end{equation}
In many of our formulas the following rescaled $\wp$-function appears:
\begin{equation}
\label{wpdefn}
\widehat\wp(z|\tau) := \frac{\wp(z,\tau)}{4\pi^2}=
x\frac{\p}{\p x}\Big( x\frac{\p}{\p x}\log\theta_1(z|\tau) \Big)-\frac{1}{12}E_2(\tau)\,.
\end{equation}
Under S duality, this transforms as 
\begin{align}
\widehat\wp(z|\tau)\rightarrow\tau^2\,\widehat\wp(z|\tau)\,.
\end{align}
A few terms that appear in the expansion of $\widehat\wp(z|\tau)$ are as follows :
\begin{align}
\label{wptexp}
\widehat\wp(z|\tau)=-\frac{1}{12}-(x+2x^2+3x^3+4x^4)+q^2\left(2-\frac{1}{x}\right)+\ldots
\end{align}
There are also the $\wp$ functions with arguments shifted by half-periods $z\rightarrow z+ \omega_i$,  
where 
\begin{align}
\label{halfperiods}
\omega_1=\frac{1}{2},\quad \omega_2=\frac{\tau}{2},\quad \omega_3=\frac{\tau+1}{2}
\end{align}
On $x$ these correspond to the following transformations respectively,
\begin{align}
\label{halfperiodtransformation}
 x\rightarrow -x,\quad\ x\rightarrow qx,\quad x\rightarrow -qx
\end{align}
The expansions for $ \widehat\wp\left(z+\omega_i|q\right)$ are easily obtained by performing \eqref{halfperiodtransformation} in \eqref{wptexp}.
The expression for $\widehat\wp$ function evaluated at the half-periods $\omega_i$ \eqref{halfperiods} are denoted as $\widehat e_i$ and they satisfy the following relations :
\begin{align}
 \widehat e_1-\widehat e_2&=\frac{\theta_3^4}{4}\cr
 \widehat e_3-\widehat e_2&=\frac{\theta_2^4}{4}\cr
\widehat e_1-\widehat e_3&=\frac{\theta_4^4}{4}
\end{align}
\section{Verifying the resummation map}
\label{appb}
Let us now verify that \eqref{x0map} is indeed the correct map that relates the bare and the renormalized variables. We start by expressing the twisted superpotential in the massless limit as the integral of the SW differential as described in Section \ref{section3}. We substitute \eqref{Gdifffgeneral} and \eqref{quaddiff} in \eqref{GGSproposal} to get :
\begin{align}
\label{w0appendix}
w_0=\int^{x_0}\sqrt{q_0(q_0-1)\frac{\partial f_0}{\partial q_0}}\,\frac{dt}{\sqrt{t(t-q_0)(t-1)}}
\end{align}
We notice that when expressed in terms of $q$ using \eqref{renocoupling} or its expansion in \eqref{qexp} we have the following:
\begin{align}
q_0\frac{\partial f_0}{\partial q_0}&=\frac{a^2}{\theta_4^4}\cr
q_0-1&=\frac{e_3-e_1}{e_1-e_2}(q)=-\frac{\theta_4^4(q)}{\theta_3^4(q)}
\end{align}
We substitute this in \eqref{w0appendix} to get the following expression for $w_0$ :
\begin{align}
\label{w0beforemani}
w_0=\frac{ia}{\theta_3^2}\int^{x_0}\frac{dt}{\sqrt{t(t-q_0)(t-1)}}
\end{align}
Let us now look at the expression for $\widetilde w_0$ in \eqref{wotilde} expressed in terms of $z$ in \eqref{newvars} and perform some simple manipulations :
\begin{align}
\label{w0mani}
\widetilde w_0=2\pi i a\int^z dz=2\pi i a\int^{z(x_0)}\,\frac{dx_0}{\frac{dx_0}{dz}}=2\pi ia\,\pi^2\theta_3^4\int^{z(x_0)}\frac{dx_0}{\wp'}\,.
\end{align} 
In the final equality we have used the map \eqref{x0map}.
The Weierstra\ss~$\wp$-function satisfies the differential equation :
\begin{align}
\wp'^2=4(\wp-e_1)(\wp-e_2)(\wp-e_3)
\end{align}
which can be expressed as 
\begin{align}
\wp'=2\pi^3\theta_3^6\sqrt{x_0(x_0-q_0)(x_0-1)}
\end{align}
We substitute the above in \eqref{w0mani} and arrive exactly at \eqref{w0beforemani}, thus confirming our $x_0$ vs $x$ map in \eqref{x0map}.

\bibliographystyle{JHEP}

\end{document}